\documentclass[aps,prc,twocolumn,showpacs,preprintnumbers,amsmath,amssymb]{revtex4-1}

\usepackage{amsmath}
\usepackage{graphicx}
\usepackage{dcolumn}
\usepackage{bm}

\begin{document}


\title{Collective flow effects on charge balance correlations and local parity-violation observables in $\sqrt{s_{NN}}=2.76$ TeV Pb+Pb collisions at the LHC}

\author{Y.~Hori$^{\mbox{1}}$}
\email{yhori@cns.s.u-tokyo.ac.jp}

\author{T.~Gunji$^{\mbox{1}}$}
\author{H.~Hamagaki$^{\mbox{1}}$}
\author{S.~Schlichting$^{\mbox{2}}$}

\affiliation{%
\vspace{0.2cm}
  $^{\mbox{1}}$Center for Nuclear Study, Graduate School of Science, the University
  of Tokyo, 7-3-1 Hongo, Bunkyo-ku, Tokyo, 113-0033, Japan}
\affiliation{%
\vspace{0.2cm}
  $^{\mbox{2}}$Universit$\ddot{a}$t Heidelberg, Institut f$\ddot{u}$r Theoretische
  Physik, Philosophenweg 16, 69120 Heidelberg, Germany
}

\date{\today} 

\begin{abstract}
We study the effects of collective flow on charge dependent azimuthal correlations at LHC energies. We propose a series of correlations as a signature of the combined effects of azimuthal collective flow and local charge conservation and perform an analysis within a statistical freeze-out model. We find that present LHC measurements of charge dependent azimuthal correlations are consistent with local charge conservation on the kinetic freeze-out surface. In view of experimental searches for signatures of the chiral-magnetic effect, we provide an alternative explanation of the charge dependence of the observed signal and propose additional measurements to disentangle the effects. 
\end{abstract}


\maketitle

\section{Introduction}
In relativistic heavy-ion collisions a large number of charged particles is produced throughout the dynamical evolution of the system. While this is readily evident from measurements of the hadronic spectra in the final state of the evolution, the spectra provide only little insight into how and when the production of charged particles occurs. In contrast measurements of charged particle correlations contain additional information about the evolution of the system and several observables have been proposed to study in particular the chemical evolution of the system \cite{bib:BF1,bib:BF2,bib:ScottCB1,bib:ScottCB2}. In this context it is important to realize, that charged particle production is subject to microscopic conservation laws, which require local production of charge anti-charge pairs in coordinate space. The strong collective expansion of the system transforms these correlations from coordinate space to momentum space, while at the same time diffusive interactions with the medium reduce the correlations \cite{bib:BF1,bib:BF2}. Charge balance correlations therefore contain important information on the creation mechanisms of charged particles and their subsequent evolution.\\
\\
In the literature it is usually assumed that charge production happens at two different stages of the evolution \cite{bib:BF1,bib:BF2,bib:ScottCB1,bib:ScottCB2}. The first wave of production is expected to occur during the first fm/$c$ of the collision by initial hard scatterings and during the thermalization process to form the quark-gluon plasma phase . After a collective expansion of the system, a second stage of charge production is expected during the transition from the deconfined QGP phase to hadronic matter around $5-10$ fm$/c$. The quark anti-quark pairs produced early in the collision are subject to diffusion over a large period of time and one expects the correlation of balancing partners to be significantly reduced in the final state. In contrast charge anti-charge pairs which are produced at times close to the kinetic freeze-out of the system exhibit diffusion only over a short period of time, and one expects a strong final state correlation. Charge balance correlations are therefore sensitive to the production time of charges as well as the transport properties of the medium. While it is an ongoing effort to disentangle the different effects \cite{bib:ScottCB1,bib:ScottCB2,bib:QBF}, we will study here to what extent experimental observations at the LHC are consistent with highly localized charge conservation at kinetic freeze-out. We will compare results from a statistical freeze-out model which takes into account local charge conservation at freeze-out \cite{bib:soeren1,bib:soeren2} to present LHC data \cite{bib:ALICE1} and provide predictions in cases where observables have not yet been measured.\\
\\
The observable employed in experimental studies of balancing charge correlations is the charge balance function
\begin{eqnarray}
\label{eq:bf}
B_{c\overline{c}}(p_{\beta}|p_{\alpha}) &&= \frac{ N_{c\overline{c}}(p_{\beta}|p_{\alpha}) - N_{cc}(p_{\beta}|p_{\alpha}) }{ dM/dp_{\alpha} } \nonumber \\
&+&  \frac{ N_{\overline{c}c}(p_{\beta}|p_{\alpha}) - N_{\overline{c}\overline{c}}(p_{\beta}|p_{\alpha}) }{ dM/dp_{\alpha} } 
\end{eqnarray}
which aims to identify balancing partner charges on a statistical basis \cite{bib:BF1,bib:BF2}. Here $dM/dp_{\alpha}$ denotes the differential charged particle multiplicity and $N_{c\overline{c}}(p_\beta|p_\alpha)$ is the number of particle pairs where one particle has charge $c$ and momentum $p_\alpha$ and the other has charge $\overline{c}$ and momentum $p_{\beta}$. The balance function $B_{c\overline{c}}(p_{\alpha}, p_{\beta})$ describes the conditional probability to observe a particle with charge $\overline{c}$ and momentum $p_\beta$, given the observation of a particle with opposite charge $c$ and momentum  $p_{\alpha}$. Instead of considering the six-dimensional correlation function in Eq. (\ref{eq:bf}), previous studies have focussed on the integrated correlation functions $B(\Delta\phi,\phi)$ and $B(\Delta\eta)$  which quantify the separation of balancing charges in azimuthal angle $\Delta\phi$  and relative pseudorapidity $\Delta\eta$ \cite{bib:BF1,bib:BF2,bib:ScottCB1,bib:ScottCB2,bib:soeren1,bib:soeren2,bib:STARBF,bib:QBF,bib:BozekFlow,bib:BozekHydro,bib:Hui}. As a consequence of the anisotropy of the system in the transverse plane the balance function $B(\Delta\phi,\phi)$ is sensitive also to the angle $\phi$ of the pair with respect to the reaction plane \cite{bib:soeren1,bib:soeren2,bib:Hui}. This behavior also constitutes a 'background effect' in experimental searches for the chiral magnetic effect \cite{bib:CME} and has triggered and intensive discussion on the interpretation of experimental results \cite{bib:soeren1,bib:soeren2,bib:Hui,bib:STARparity,bib:ScottParity,bib:Adam1,bib:Adam2,bib:AdamSuppr,bib:AMPT}. \\
\\
In this paper we focus on the balance function $B(\Delta\phi,\phi)$ and study the dependence on the reaction plane angle $\phi$ at LHC energies. We employ a statistical freeze-out model which takes into account local charge conservation at kinetic freeze-out, in order to estimate the combined effects of collective flow and local charge conservation on charge dependent correlation functions. In addition to the charge balance function, we also consider mixed harmonic azimuthal correlations, $\langle {\rm
  cos}(m(\phi_{\alpha}-\phi_{\beta}) - n (\phi_{\beta}-\Psi_{k})) \rangle$, which correspond to different moments of the balance
function with respect to the $k$-th order reaction plane $\Psi_{k}$.  
In this way we can investigate not only the separation width of the
balancing charge but also its variation with respect to the event plane, which is sensitive to the flow properties of charged particle pairs.  One of the correlations we investigate, $\langle {\rm  cos}(\phi_{\alpha}+\phi_{\beta}-2\Psi_{RP}) \rangle$, has originally been
proposed in the search of experimental signatures of the chiral magnetic effect (CME) \cite{bib:sergei}.  In measurements of the STAR and ALICE experiments \cite{bib:STARparity,bib:ALICE1}, it was found that this correlation
shows a strong charge dependence which has been argued to be consistent to the
CME \cite{bib:STARparity,bib:ALICE1}. However the combined effects of local charge conservation and elliptic flow have been shown to account for the charge dependence of the signal observed at RHIC \cite{bib:soeren1,bib:soeren2,bib:Hui}. We will present results, which show that this statement holds also for the LHC measurement and we propose to systematically measure a series of
correlations to disentangle the different effects.

\section{Freeze-Out model}
\begin{figure}[t]
  \centering

\begin{tabular}{|c|c|c|c|c|c|c|}
\hline
centrality & T  & n     & $\beta_{0}$ & $\beta_{2}$ & $\beta_{4}$  & $\epsilon$\\
\hline \hline
5-10\% &  0.086 & 1.10  & 0.930     & 0.00831      &  0.0015       &  0.0961\\
10-20\% & 0.085 & 1.17  & 0.935     & 0.0129       & 0.00203       & 0.130\\
20-30\% & 0.083 & 1.24  & 0.938     & 0.0154       & 0.00255       & 0.172\\
30-40\% & 0.080 & 1.35  & 0.940     & 0.0160       & 0.00285       & 0.204 \\
40-50\% & 0.080 & 1.46  & 0.943     & 0.0151       & 0.00333       & 0.224\\
50-60\% & 0.079 & 1.50  & 0.944     & 0.0136       & 0.00316       & 0.226\\
\hline
\end{tabular}

  \includegraphics[width=0.45\textwidth]{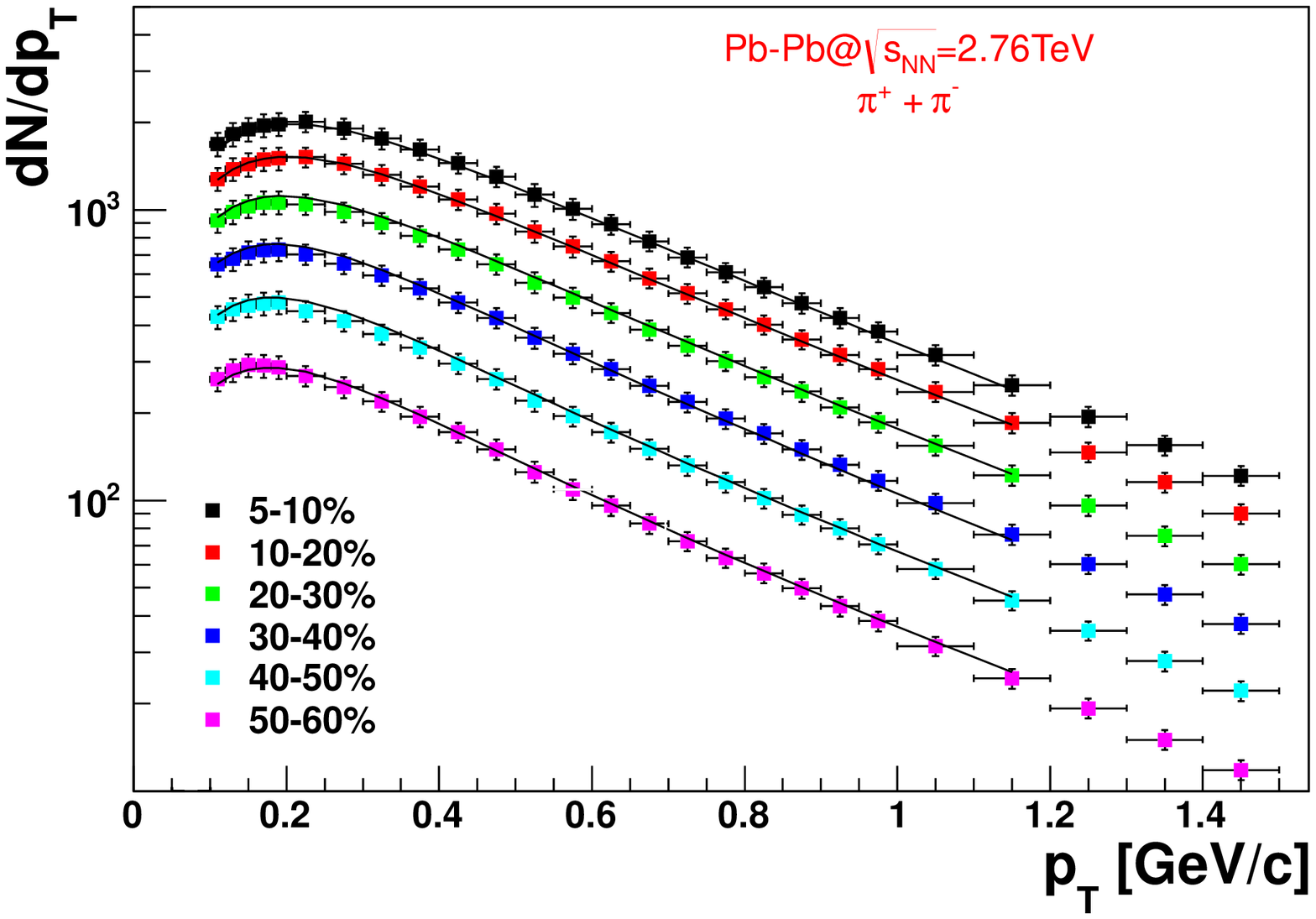}
  \includegraphics[width=0.45\textwidth]{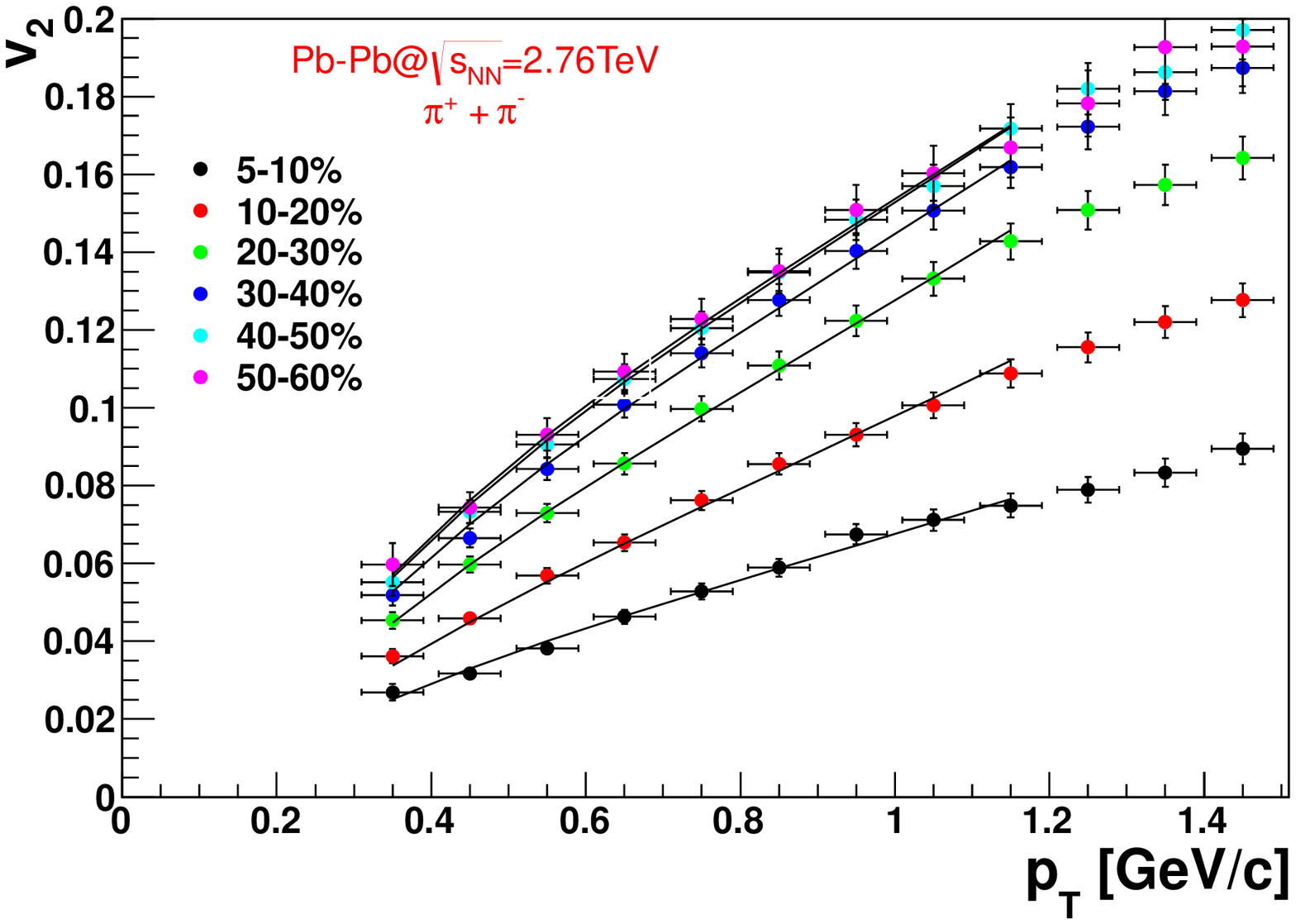}
  \caption{Charged particle yield $dN/dp_{T}$ and elliptic flow $v_2(p_T)$ from the ALICE
  collaboration \cite{bib:ALICE2,bib:ALICE3} and the blast wave model for six different centrality classes. The parameters of the blast wave model are summarized in the above table.}
  \label{fig:BWsp}
\end{figure}

We use the freeze-out model developed in~\cite{bib:soeren1, bib:soeren2} in order to calculate charge
dependent correlations in $\sqrt{s_{NN}}=2.76$ TeV Pb+Pb collisions at the LHC. The model is based on conventional blast wave models, where in addition local charge conservation on the kinetic freeze-out surface is taken into account \cite{bib:soeren1,bib:soeren2}. Here we use a filled ellipse version of the blast wave model, which is a simple parameterization of the freeze-out configuration~\cite{bib:BW}. The model parameters of the blast-wave model are the kinetic freeze-out temperature $T_{kin}$, the flow parameters $n$, $\beta_{0}$, $\beta_{2}$, $\beta_{4}$, the eccentricity of the freeze-out surface $\epsilon$ and the chemical freeze-out temperature $T_{ch}$. The transverse collective flow velocity is parameterized as
\begin{eqnarray}
\label{eq:tflow}
\beta(r, \phi_{b}) = \tilde{r}^{n}(\phi_{b})(\beta_{0} &+& \beta_{2}{\rm cos}(2\phi_{b}) + \beta_{4}{\rm cos}(4\phi_{b})) 
\end{eqnarray}
The model parameters are entirely determined by single particle spectra and are chosen to reproduce LHC results of  $v_{2}$, $v_{4}$ and $p_{T}$ spectra from the ALICE collaboration \cite{bib:ALICE2,bib:ALICE3}. The parameters are summarized in the above table and the model is compared to the ALICE data in Fig. \ref{fig:BWsp}. In addition to the single particle parameters we consider the parameter $\sigma_{\phi}$, which control the separation of balancing charges on the freeze-out surface \cite{bib:soeren2}. We will mostly consider the case $\sigma_{\phi}=0$, where balancing partner charges are emitted from the same point on the freeze-out surface. This amounts to perfectly local charge conservation and gives rise to the strongest possible correlations. In some cases we vary the parameter $\sigma_{\phi}$ to study the sensitivity of our results.
The implementation of local charge conservation is identical to Refs. \cite{bib:soeren1,bib:soeren2} and proceeds as follows: In  the first step of calculation we create arrays of particles according to a canonical ensemble with fixed volume (100 fm$^{3}$) and chemical freeze-out temperature $T_{ch}$ = 148 MeV~\cite{bib:ALICE3}, such that electric charge, strangeness and baryon number all sum to zero. We then randomly assign a point on the freeze-out surface to each array, which determines the collective flow velocity. Finally the momentum of each particle is assigned according to a thermal distribution with the collective velocity of the array. Charge dependent correlations can be calculated separately for each array as particles from different arrays are assumed to be uncorrelated in our model. We note that in reality, charge dependent correlations are sensitive also to the correlations induced 
by final-state interactions and identical particle interference \cite{bib:BFdist}. When incorporated in the model, these have been shown to explain distortions of the balance function at RHIC \cite{bib:soeren2,bib:BFdist}. However, for our purpose of estimating the magnitude of the local charge conservation signal, we expect these effects to be less relevant and we will neglect them in the following.
%
\section{Charge dependent azimuthal correlations at LHC energy}
%
\begin{figure}[t]
  \centering
  \includegraphics[width=0.45\textwidth]{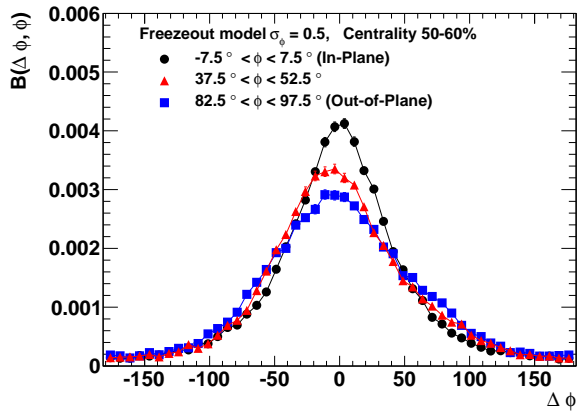}
  \caption{The balance function with respect to the reaction plane calculated from the freeze-out model. One observes that the balance funtion is narrower for in-plane pairs as compared to out-of plane pairs. The balance function at intermediate angles shows an asymmetry in $\Delta\phi$ towards the reaction plane.}
  \label{fig:Bphi}
\end{figure}
\begin{figure}[t]
  \centering
  \includegraphics[width=0.45\textwidth]{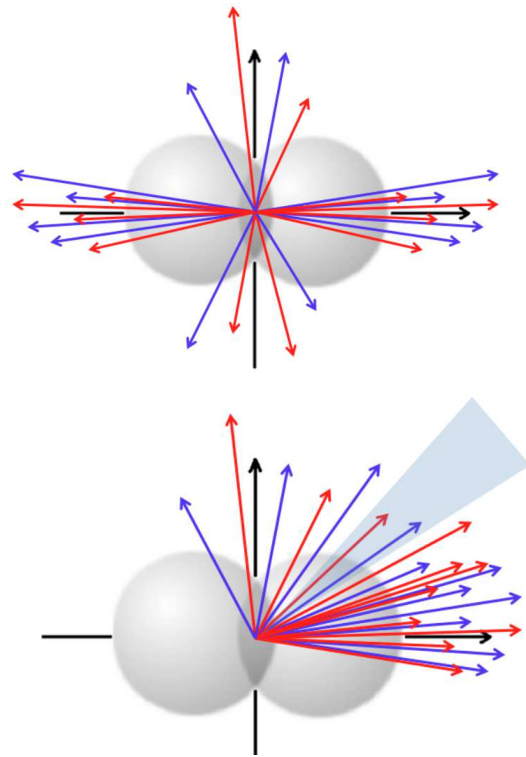}
  \caption{Illustration of the combined effects of collective flow and local charge conservation. The top panel shows how in-plane pairs are more tightly correlated in azimuthal angle as compared to out of plane pairs, due to larger collective velocities in-plane. To lower panel illustrates that at intermediate angles the balancing partner charge is more likely to be found towards the in-plane direction.}
  \label{fig:Cartoon}
\end{figure}

Here we present results for charge dependent azimuthal correlations from the blast wave model described in the previous section. In order to study the combined effects of collective flow and local charge conservation, we first investigate the charge balance function in azimuthal angle and with respect to the reaction plane
\begin{eqnarray}
B(\Delta \phi, \phi) &\equiv& \frac{1}{dM/d\phi}
\int dp_{\alpha}
\frac{dM}{dp_{\alpha}}dp_{\beta}~B(p_{\alpha}|p_{\beta})\nonumber \\
&& \delta(\Delta
\phi -(\phi_{\alpha} - \phi_{\beta}))\delta(\phi -\phi_{\beta})\;.
\end{eqnarray}
Here $\phi$ is the azimuthal angle of the leading particle with respect to the reaction plane and  $\Delta\phi$ is the relative angle of the pair. The results are shown in Fig. \ref{fig:Bphi}, where we present the balance function $B(\Delta\phi,\phi)$ as a function of $\Delta\phi$ for $\phi\simeq0^{\circ},45^{\circ}$ and $90^{\circ}$ and we use the same kinematic cuts ($|\eta|<0.8,~0.2\text{GeV}/c p_T < 5 \text{GeV}/c$) as the ALICE collaboration. From Fig. \ref{fig:Bphi} one observes that the balance function in-plane ($\phi\simeq0^{\circ}$) is significantly narrower as compared to the balance function out-of-plane ($\phi\simeq90^{\circ}$). This is a consequence of the larger collective velocities in-plane, which lead to a focussing in azimuthal angle as illustrated in the top panel of Fig. \ref{fig:Cartoon}. At intermediate angles ($\phi\simeq45^{\circ}$) the balance function shows an asymmetry towards the in-plane direction. This can be attributed to the fact that the balancing partner charge is statistically more likely to be found in-plane as compared to out-of-plane as illustrated in the lower panel of Fig. \ref{fig:Cartoon}.\\
\\
\begin{figure}[t]
  \centering
  \includegraphics[width=0.5\textwidth]{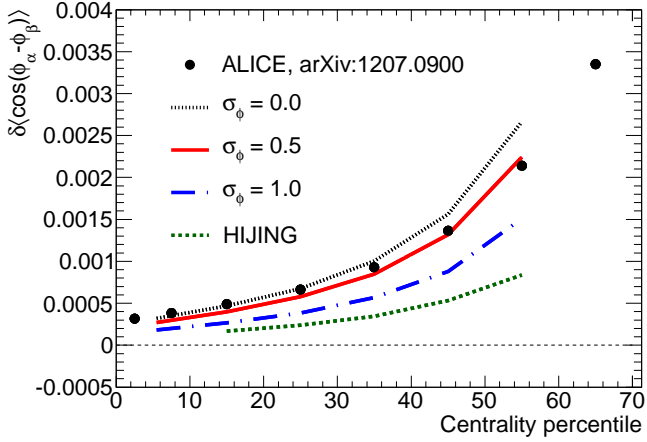}
  \caption{The centrality dependence of the integrated correlation
  $\delta \langle {\rm cos} (\Delta \phi) \rangle$ calculated by the
  freeze-out model. The different values of
  $\sigma_{\phi}$ correspond to different separations of balancing partners on the freeze-out surface. Black circles are the results from the ALICE
  collaboration. The results from HIJING simulations (green dot line), which do not include local charge conservation clearly underestimate the observed signal.}
  \label{fig:width}
\end{figure}
\begin{figure}[t]
  \centering
  \includegraphics[width=0.5\textwidth]{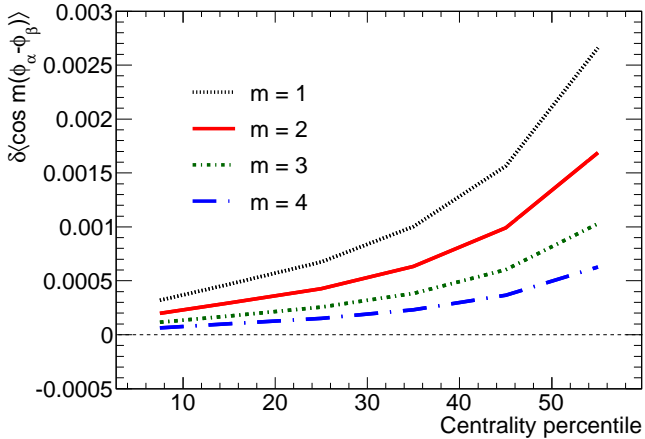}
  \caption{The centrality dependence of the integrated correlation
  $\delta \langle {\rm cos} (m\Delta \phi) \rangle$ calculated by the
  freeze-out model with $\sigma_{\phi}=0.0$. The correlators quantify higher harmonic moments of the charge balance function.}
  \label{fig:widthn}
\end{figure}
The modulation of the balance function in the azimuthal angle of the pair $\phi$, can be quantified in terms of harmonic moments of the balance function. In order to establish this relation we introduce the set of correlators
\begin{eqnarray} 
C_{\alpha\beta}(n,m;k) \equiv \frac{\sum_{\alpha,\beta}^{\beta\neq\alpha}{\rm cos} \Bigl(m(\phi_{\alpha}-\phi_{\beta}) - n (\phi_{\beta}-\Psi_{k})\Bigr)}{M_{\alpha}M_{\beta}}\;,  \nonumber \\
\end{eqnarray}
and we will only consider the charge dependent part
\begin{eqnarray}
\delta C\equiv\frac{1}{2}\left[C_{+-}+ C_{-+}-C_{++}- C_{--}\right]\;. 
\end{eqnarray}
Here $\psi_k$ denotes the $k$-th order event plane. Since in the statistical model the event planes of different orders coincide, we will restrict ourselves to $k=2$ in the following. However in the presence of flow fluctuations this is no longer the case and it would be interesting to also study the behavior in event-by-event simulations. The correlators $\delta C(n,m;2)$ are related to the charge balance function by
\begin{eqnarray}
\frac{\langle M^2 \delta C(n,m;2)\rangle}{\langle M \rangle}&=&\frac{2}{\langle M \rangle}\int~d\phi~d\Delta\phi~\langle \frac{dM}{d\phi} \rangle B(\phi,\Delta\phi) \nonumber \\
&&\qquad\quad\times~{\rm cos}(m\Delta\phi+n\phi)\;,
\end{eqnarray}
such that $C(n,m;2)$ quantifies the $n$-th harmonic modulation of the $m$-th harmonic moments of balance function. The zeroth moment $\delta C(0,1;2)=\delta\langle {\rm cos}(\Delta \phi)\rangle$ quantifies the overall width of the charge balance function and is shown in Fig. \ref{fig:width}. The higher (unmodulated) moments $\delta C(0,m;2)=\delta\langle {\rm cos}(m\Delta \phi)\rangle$ are shown in Fig. \ref{fig:widthn}. In order to compare our model results to ALICE data, we account for the detector efficiency and acceptance by using the same kinematic cuts as the ALICE experiment and rescaling our data to reproduce the experimental multiplicities and the overall normalization \cite{bib:ALICE4}
\begin{eqnarray}
Z=(N_{+-}-N_{++})/N_{+}+(N_{-+}-N_{--})/N_{-}\;.
\end{eqnarray}
One observes that for central and mid-central collisions the ALICE results are compatible with perfectly local charge conservation on the freeze-out surface. For more peripheral collisions we find that local charge conservation over-estimates the signal. However if we reduce the degree of correlation by taking into account a non-zero separation $\sigma_{\phi}$ of balancing charges on the freeze-out surface, we find that the model results are compatible with the ALICE data. We note that final-state interactions and identical particle interference might further reduce the correlation, such that charge conservation is more local than suggested by Fig. \ref{fig:width}. This effect can be investigated by measurements of the balance function $B(\phi)$ as well as measurements of the higher moments $\delta C(0,m;2)$. For comparison we also show results from HIJING simulations, which do not account for local charge conservation and clearly underestimate the signal.\\
\\
\begin{figure}[t]
  \centering
  \includegraphics[width=0.47\textwidth]{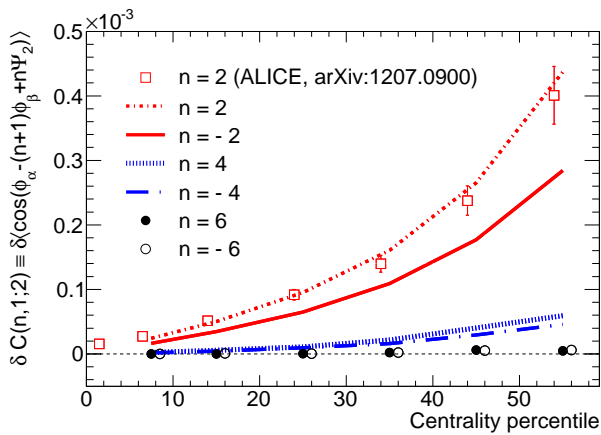}
  \caption{The centrality dependence of the integrated correlation
  $\delta C(n,1;2)$ calculated by the freeze-out model. The correlation
  $\delta C(2,1;2)$ has been measured by the ALICE collaboration in the context of local strong parity violation. We find that local charge conservation provides an alternative explanation of the charge dependent signal.}
  \label{fig:CDC}
\end{figure}
The modulated moments $n>0$ generically feature three different contributions, which can be written as \cite{bib:ScottParity,bib:soeren1,bib:soeren2}
\begin{eqnarray}
\delta C(n,m;2)=v_{n}C_{m}+v^{(m)}_{n,c}-v^{(m)}_{n,s} \;,
\end{eqnarray}
where we introduced the flow coefficients $v_{n}$ and the moments of the balance function $C_{m}$ as
\begin{eqnarray}
v_{n}&=&\langle {\rm cos}(n\phi)\rangle \;, \\
C_m&=&\delta\langle {\rm cos}(m\Delta\phi)\rangle\;.
\end{eqnarray}
The contribution $v_{n}C_{m}$ is a combination of the harmonic modulation of the single particle spectra and the charge balance width. For instance for the second harmonic moment $\delta C(2,1;2)$, the contribution $v_2C_1$ originates from higher particle multiplicites in-plane as compared to out-of-plane \cite{bib:ScottParity,bib:soeren1,bib:soeren2}. The other contributions take the form
\begin{eqnarray}
v^{(m)}_{n,c}&=&\langle {\rm cos}(m\Delta\phi){\rm cos}(n\phi)\rangle-v_{|n|}C_m\;, \\
v^{(m)}_{n,s}&=&\langle {\rm sin}(m\Delta\phi){\rm sin}(n\phi)\rangle\;.
\end{eqnarray}
which are associated to the modulation of the charge balance function in the azimuthal angle of the leading particle. Concerning the second harmonic moment $\delta C(2,1;2)$ one finds that the correlator $v_{2,c}$ describes the modulation of the width of the balance function, whereas $v_{2,s}$ is negative and originates from the asymmetry of the balance function at intermediate angles \cite{bib:ScottParity,bib:soeren1,bib:soeren2}. These features can clearly be observed from Fig. \ref{fig:Bphi} and similarly one can identify the contributions for $n=4,6,...$ associated to higher order flow harmonics. The blast-wave results for the correlation function $\delta C(n,1;2)$ are presented in Fig. \ref{fig:CDC} as a function of centrality. The correlator $\delta C(2,1;2)$ has originally been proposed as a signature of the chiral magnetic effect \cite{bib:sergei} and been measured by the ALICE collaboration \cite{bib:ALICE1}. The results in Fig. \ref{fig:CDC} show that local charge conservation already accounts for the entire charge dependence of the observed signal. We emphasize that this observation is in complete analogy to the situation at RHIC \cite{bib:soeren1,bib:soeren2,bib:Hui}. From Fig. \ref{fig:CDC} one also observes a large signal also for the $n=\pm4$ harmonic components, which originates from the quadrangular anisotropy and we suggest to use this correlator to verify the origin of the correlation. On the other hand correlations with $n=\pm6$ are highly suppressed because $v_{6}$\{$\Psi_{2}$\} is very small. In the context of the experimental search for signatures of the chiral magnetic effect, it might therefore be interesting to investigate the higher moments ($|n|>4$), where the background contribution from local charge conservation is significantly reduced.

\section{Summary and Discussion}
In this paper we studied the combined effects of collective flow and local charge conservation on charge dependent correlation functions at LHC energies. We employed the blast wave model from Ref. \cite{bib:soeren1,bib:soeren2} and performed a fit to single particle spectra to extract the blast wave parameters. Within this model we studied charge balance correlations in the transverse plane. Similar to the results at RHIC energies \cite{bib:soeren1,bib:soeren2,bib:Hui}, we find that collective flow leads to a modulation of charge balance correlations with respect to the reaction plane. This can be quantified in terms of the moments $\delta C(n,m;k)$ of the balance function which we propose to measure. When comparing our model results to LHC measurements in the context of searches for the chiral magnetic effect \cite{bib:ALICE1}, we find that the charge dependence observed in the ALICE data can be fully explained with local charge conservation at kinetic freeze-out. We also provided predictions for additional correlators to test this alternative explanation of the signal. We emphasize that the notion of local charge conservation at kinetic freeze-out is an interesting observation by itself. This is because the correlations of balancing charges in the final state are directly related to the chemical evolution of the plasma as well as the transport properties of the medium. The key challenge is to identify the origin of local charge conservation at freeze-out which can be explained by small in-medium diffusion of quarks produced in the early stage of the evolution  as well as by a different production mechanism where charged particles are predominantly produced at late times. In this context it was pointed out that balance functions of identified charged particles provide a promising tool to disentangle hadronic and partonic contributions \cite{bib:ScottCB1, bib:ScottCB2,bib:QBF} and one should also take into account the recent results of direct photon measurements, which point to strong collective behavior early in the collision \cite{bib:PHENIX}. Finally it would be interesting to study the effects of local charge conservation in event-by-event simulations and possibly also within a microscopic approach. First results in this direction have been presented in \cite{bib:BozekHydro,bib:AMPT} and we expect more studies in the future.\\
\\  
\textit{Acknowledgement:} The authors like to thank A.~Bzdak and S.~Pratt for insightful discussion. Y.~Hori was the JPSP Research Fellow and this work was supported in part by JSPS grant No.~10J08478 (Y.~Hori) and JSPS program No.~12601 (H.~Hamagaki).


\begin{thebibliography}{10}
\bibitem{bib:ALICE4}
  Experimental multiplicities and $(N_{+-}-N_{++})/N_{+} + (N_{-+}-N_{--})/N_{-}$ were generously provided by the ALICE collaboration.
\end{thebibliography}

\end{document}